\documentclass{article}%
\usepackage{graphicx}
\usepackage{amsmath}
\usepackage{amsfonts}
\usepackage{amssymb}%
\providecommand{\U}[1]{\protect\rule{.1in}{.1in}}

\begin{document}

\author{Antony Valentini\\Augustus College}

\begin{center}
{\LARGE Instability of quantum equilibrium in Bohm's dynamics}

\bigskip

\bigskip

\bigskip

\bigskip\ 

Samuel Colin$^{1,2}$, Antony Valentini$^{1}$

{\small \ }

$^{1}$\textit{Department of Physics and Astronomy,}

\textit{Clemson University, Kinard Laboratory,}

\textit{Clemson, SC 29634-0978, USA.}

{\small \ }

$^{2}$\textit{Centre for Quantum Dynamics,}

\textit{Griffith University,}

\textit{Brisbane, QLD 4111, Australia}.

\bigskip

\bigskip

\bigskip

\bigskip
\end{center}

We consider Bohm's second-order dynamics for arbitrary initial conditions in
phase space. In principle Bohm's dynamics allows for `extended'
nonequilibrium, with initial momenta not equal to the gradient of phase of the
wave function (as well as initial positions whose distribution departs from
the Born rule). We show that extended nonequilibrium does not relax in general
and is in fact unstable. This is in sharp contrast with de Broglie's
first-order dynamics, for which non-standard momenta are not allowed and which
shows an efficient relaxation to the Born rule for positions. On this basis we
argue that, while de Broglie's dynamics is a tenable physical theory, Bohm's
dynamics is not. In a world governed by Bohm's dynamics there would be no
reason to expect to see an effective quantum theory today (even
approximately), in contradiction with observation.

\bigskip

\bigskip

\bigskip

\bigskip

\bigskip

\bigskip

\bigskip

\bigskip

\bigskip

\bigskip

\bigskip

\bigskip

\bigskip

\bigskip

\bigskip

\bigskip

\bigskip

\bigskip

\bigskip

\bigskip

\bigskip

\bigskip

\bigskip

\bigskip

\bigskip

\bigskip

\section{Introduction}

In 1927 de Broglie proposed a new form of dynamics for a many-body system (de
Broglie 1928). For $N$ non-relativistic and spinless particles, with
configuration $q(t)=(\mathbf{x}_{1}(t),\mathbf{x}_{2}(t),...,\mathbf{x}%
_{N}(t))$, the particle velocities at time $t$ are given by de Broglie's
guidance equation%
\begin{equation}
\frac{d\mathbf{x}_{i}}{dt}=\frac{\mathbf{\nabla}_{i}S}{m_{i}} \label{geqn}%
\end{equation}
(with masses $m_{i}$ and $i=1,2,...,N$), where $S$ is the phase of a complex
wave $\Psi(q,t)$ in configuration space that satisfies the Schr\"{o}dinger
equation%
\begin{equation}
i\frac{\partial\Psi}{\partial t}=\sum_{i=1}^{N}-\frac{1}{2m_{i}}\nabla_{i}%
^{2}\Psi+V\Psi\label{Seqn}%
\end{equation}
in the presence of an external classical potential $V$ (where $\hslash=1$ and
$\Psi=\left\vert \Psi\right\vert e^{iS}$). De Broglie called this theory
`pilot-wave theory', and he presented it at the fifth Solvay conference as a
theory of microscopic quantum systems (Bacciagaluppi and Valentini 2009,
Valentini 2009)\textbf{.}

As is now well known, the empirical predictions of quantum mechanics may be
derived from de Broglie's dynamics -- defined by (\ref{geqn}), (\ref{Seqn}) --
provided it is assumed that an ensemble of systems with initial wave function
$\Psi(q,0)$ has initial configurations $q(0)$ that are distributed according
to the Born rule, with a probability density%
\begin{equation}
P(q,0)=|\Psi(q,0)|^{2} \label{Born}%
\end{equation}
in configuration space at $t=0$. This was shown fully by Bohm in 1952 (Bohm
1952a,b). A key point in the derivation is to apply the dynamics to the
apparatus, as well as to the microscopic system, and to show that the
distribution of apparatus readings (over an ensemble of experiments) agrees
with quantum theory.

It is an elementary consequence of (\ref{geqn}), (\ref{Seqn}) that the
Born-rule distribution $P=|\Psi|^{2}$ is preserved in time: if it holds at
$t=0$, it will hold at all times. To see this note first that, because each
element of the ensemble moves with velocity\newline%
\[
\dot{q}=(\mathbf{\dot{x}}_{1},\mathbf{\dot{x}}_{2},...,\mathbf{\dot{x}}%
_{N})=(\mathbf{\nabla}_{1}S/m_{1},\mathbf{\nabla}_{2}S/m_{2}%
,...,\mathbf{\nabla}_{N}S/m_{N})\ ,
\]
the ensemble distribution $P(q,t)$ necessarily obeys the continuity equation%
\[
\frac{\partial P}{\partial t}+\nabla_{q}\cdot(P\dot{q})=0
\]
(where $\nabla_{q}=(\mathbf{\nabla}_{1},\mathbf{\nabla}_{2},...,\mathbf{\nabla
}_{N})$). Furthermore, as is well known, the Schr\"{o}dinger equation
(\ref{Seqn}) implies that $|\Psi|^{2}$ obeys%
\[
\frac{\partial|\Psi|^{2}}{\partial t}+\nabla_{q}\cdot(|\Psi|^{2}\dot{q})=0\ ,
\]
which is just the same continuity equation (with the same velocity field
$\dot{q}$). Thus, $P$ and $|\Psi|^{2}$ evolve according to the same partial
differential equation, and so the same initial conditions for $P$ and
$|\Psi|^{2}$ will yield the same time evolution. The distribution
$P=|\Psi|^{2}$ is therefore an equilibrium distribution, often referred to as
`quantum equilibrium'.

It has become apparent that, at least in principle, de Broglie's dynamics
contains a physics that is much wider than quantum physics, with possible
`nonequilibrium' ensemble distributions $P\neq|\Psi|^{2}$ that violate the
usual Born rule (Valentini 1991a,b, 1992, 1996, 2001, 2002, 2007, 2008a, 2009,
2010; Pearle and Valentini 2006). For there is a clear conceptual distinction
between the laws of motion (\ref{geqn}), (\ref{Seqn}) for a single system on
the one hand, and the assumption (\ref{Born}) about the distribution of
initial conditions on the other hand. In a deterministic dynamics, initial
conditions are in principle arbitrary and cannot be regarded as laws.
Therefore, if de Broglie's pilot-wave theory is taken seriously it must be
admitted that departures from the Born rule (\ref{Born}) are in principle
possible -- just as departures from thermal equilibrium are obviously possible
in classical dynamics.

It has been shown that non-Born rule distributions in pilot-wave theory can
give rise to a wealth of new phenomena. These include nonlocal signalling
(Valentini 1991b) -- which suggests that the theory contains an underlying
preferred foliation of spacetime (Valentini 2008a) -- and `subquantum'
measurements that violate the uncertainty principle and other standard quantum
constraints (Valentini 2002; Pearle and Valentini 2006). On this view, quantum
physics is a special equilibrium case of a much wider nonequilibrium physics.

As one might expect, given the analogy with thermal equilibrium, it is found
that initial nonequilibrium states relax to equilibrium -- on a coarse-grained
level, provided the initial state contains no fine-grained microstructure
(Valentini 1991a, 1992, 2001; Valentini and Westman 2005; Efthymiopoulos and
Contopoulos 2006; Bennett 2010; Towler, Russell and Valentini 2012; Colin
2012).\footnote{The latter proviso is analogous to that required in the
classical statistical mechanics of an isolated system (Davies 1977). An
assumption about initial conditions is of course required, in any
time-reversal invariant theory, for relaxation to occur. For a full discussion
see Valentini (1992, 1996, 2001) and Valentini and Westman (2005).} In
particular, relaxation has been found to occur for wave functions that are
superpositions of different energy eigenvalues. Because all the systems we
have access to have had a long and violent astrophysical history, there has
been plenty of opportunity for such relaxation to take place. Therefore, if
our world is governed by de Broglie's dynamics we should expect to see
equilibrium today -- in agreement with observation, which has confirmed the
Born rule in a wide range of conditions.\textbf{ }On the other hand, in the
context of inflationary cosmology, quantum nonequilibrium at very early times
could leave an observable imprint today on the cosmic microwave background
(Valentini 2010). It has also been shown that, in certain conditions,
relaxation can be suppressed for long-wavelength field modes in the early
universe, and it is possible that low-energy relic particles could still exist
today that violate the Born rule (Valentini 2007, 2008b; Colin and Valentini
2013). Apart from these cosmological possibilities, however, if we focus on
the physics of ordinary systems in the laboratory, then according to de
Broglie's dynamics equilibrium today is to be expected.

The aim of this paper is to provide a similar analysis for Bohm's 1952
reformulation of de Broglie's 1927 dynamics.

In Bohm's 1952 papers, the dynamics was presented in a form different from
that of de Broglie. Instead of the law of motion (\ref{geqn}) for velocities,
Bohm wrote the dynamics in a Newtonian form in terms of a law of motion for
accelerations,%
\begin{equation}
m_{i}\frac{d^{2}\mathbf{x}_{i}}{dt^{2}}=-\mathbf{\nabla}_{i}(V+Q)\ ,
\label{neqn'}%
\end{equation}
with a `quantum potential'%
\begin{equation}
Q\equiv-\sum_{i=1}^{N}\frac{1}{2m_{i}}\frac{\nabla_{i}^{2}\left\vert
\Psi\right\vert }{\left\vert \Psi\right\vert }%
\end{equation}
that is generated by $\Psi$.

To derive the predictions of quantum mechanics Bohm made \textit{two}
assumptions about the initial conditions:

(i) that initial particle positions, or configurations $q(0)=(\mathbf{x}%
_{1}(0),\mathbf{x}_{2}(0),...,\mathbf{x}_{N}(0))$, are distributed according
to the Born rule (\ref{Born}), and

(ii) that initial particle momenta are restricted to the values%
\begin{equation}
\mathbf{p}_{i}(0)=\mathbf{\nabla}_{i}S(q,0) \label{pic}%
\end{equation}
(where the right-hand side of (\ref{pic}) is determined by the initial wave
function $\Psi(q,0)$ and by the initial representative point $q$ in
configuration space).

Given the initial conditions (\ref{pic}), it follows from (\ref{neqn'}) and
(\ref{Seqn}) that at all times $t$ those conditions are preserved,%
\begin{equation}
\mathbf{p}_{i}(t)=\mathbf{\nabla}_{i}S(q,t)\ . \label{pt}%
\end{equation}
To see this, note that $\mathbf{p}_{i}$ and $\mathbf{\nabla}_{i}S$ (evaluated
along a trajectory) evolve according to the same ordinary differential
equation: specifically, we have%
\[
\frac{d\mathbf{p}_{i}}{dt}=-\mathbf{\nabla}_{i}(V+Q)
\]
and also%
\[
\frac{d}{dt}(\mathbf{\nabla}_{i}S)=-\mathbf{\nabla}_{i}(V+Q)
\]
(with $d/dt=\partial/\partial t+\dot{q}\cdot\nabla_{q}$). The latter equation
follows immediately by taking the gradient of the modified Hamilton-Jacobi
equation%
\[
\frac{\partial S}{\partial t}+\sum_{i=1}^{N}\frac{(\mathbf{\nabla}_{i}S)^{2}%
}{2m_{i}}+V+Q=0\ ,
\]
which, as is well known, follows from the Schr\"{o}dinger equation
(\ref{Seqn}). Thus, the same initial conditions for $\mathbf{p}_{i}$ and
$\mathbf{\nabla}_{i}S$ necessarily yield the same time evolution (along a trajectory).

Since (\ref{pt}) is just de Broglie's original equation of motion
(\ref{geqn}), it follows that the trajectories of Bohm's dynamics are the same
as the trajectories of de Broglie's dynamics -- provided, that is, that the
initial conditions (\ref{pic}) on the momenta are assumed. It then follows, as
in de Broglie's dynamics, that the Born-rule distribution for positions
(assumed to hold at $t=0$) will hold for all $t$, and one may then demonstrate
empirical equivalence to quantum theory.

In Bohm's dynamics, (\ref{pic}) is an initial condition which may in principle
be dropped, and the same is true of the Born rule (\ref{Born}). The condition
(\ref{pic}) happens to be preserved in time by the dynamics, yielding the
condition (\ref{pt}) at later times, but (\ref{pt}) is not itself a law of
motion. In de Broglie's dynamics, in contrast, (\ref{pt}) \textit{is} the law
of motion, and there is no question of dropping (\ref{pic}), which is simply
the law of motion applied at the initial time.

De Broglie's dynamics and Bohm's dynamics are therefore quite different, not
only in form but also in substance. De Broglie's theory contains a wider
physics, of which quantum theory is only a special case. Bohm's theory
contains an \textit{even wider} physics, of which de Broglie's theory and
quantum theory are only special cases.

This difference between the two dynamical theories has deep historical roots.
The original pilot-wave dynamics was constructed by de Broglie in the years
1923--27, with the aim of unifying the physics of particles with the physics
of waves. Among other things, de Broglie argued that to explain the
diffraction of single photons -- where the particle does not touch the
diffracting screen and yet does not move in a straight line -- Newton's first
law of motion should be abandoned. The first-order guidance equation
(\ref{geqn}) or (\ref{pt}) was proposed as the fundamental law of motion of a
new, non-Newtonian dynamics. De Broglie motivated this law as a unification of
the classical variational principles of Maupertuis ($\delta\int m\mathbf{v}%
\cdot d\mathbf{x}=0$, for a particle with velocity $\mathbf{v}$) and of Fermat
($\delta\int dS=0$, for a wave with phase $S$).\footnote{For a full discussion
see Bacciagaluppi and Valentini (2009, chapter 2).} Bohm, in contrast,
rediscovered de Broglie's theory in the early 1950s but based his presentation
on the second-order, Newtonian equation of motion (\ref{neqn'}). On Bohm's
original view, the guidance equation was to be regarded as a mere constraint
on the initial momenta, a constraint that could in principle be dropped. This
was clearly stated by Bohm in 1952 (even if this point was lost in later presentations):

\begin{quotation}
The equation of motion of a particle ... is [(\ref{neqn'})]. It is in
connection with the boundary conditions appearing in the equations of motion
that we find the only fundamental difference between the $\psi$-field and
other fields ... . For in order to obtain results that are equivalent to those
of the usual interpretation of the quantum theory, we are required to restrict
the value of the initial particle momentum to [(\ref{pic})]. ... this
restriction is consistent, in the sense that if it holds initially, it will
hold for all time. ... however, ... this restriction is not inherent in the
conceptual structure. (Bohm 1952a, p. 170)
\end{quotation}

While Bohm did not consider details of what would happen if one dropped the
initial momentum constraint (\ref{pic}), he did make clear that this
constraint is not a law. Logically, therefore, if it is not a law it may in
principle be dropped. This raises a separate question: why is (\ref{pic})
satisfied in nature? Bohm understood that (\ref{pic}) is necessary to
guarantee agreement with quantum mechanics. To explain how (\ref{pic}) might
arise, Bohm (1952a, p. 179) tentatively suggested modifying the law of motion
(\ref{neqn'}) in such a way that (\ref{pic}) becomes an attractor. However,
the focus of Bohm's paper concerned what we call Bohm's dynamics, with
(\ref{neqn'}) as the equation of motion and (\ref{pic}) as an arbitrary
initial condition.

To summarise, de Broglie and Bohm proposed two quite distinct forms of
dynamics, which become equivalent only by assuming the initial condition
(\ref{pic}) on the momenta. In the context of Bohm's dynamics, if one is
unwilling to consider dropping (\ref{pic}) then one may as well use (\ref{pt})
as the law of motion -- thereby in effect abandoning Bohm's dynamics in favour
of de Broglie's. Thus, if one wishes to regard Bohm's dynamics as fundamental
then one should consider relaxing (\ref{pic}) at least in principle.

On a point of terminology, we remark that the term `Bohmian mechanics'\ is
sometimes used (misleadingly) by some workers to denote de Broglie's
first-order dynamics. To avoid confusion, throughout this paper we use the
term `Bohm's dynamics' to refer specifically to the second-order dynamics
defined by equations (\ref{Seqn}) and (\ref{neqn'}), which we distinguish
sharply from what we call `de Broglie's dynamics' -- the first-order dynamics
defined by equations (\ref{geqn}) and (\ref{Seqn}).

In this paper we shall study Bohm's dynamics with what we call `extended
nonequilibrium', that is, with initial momenta $\mathbf{p}_{i}\neq
\mathbf{\nabla}_{i}S$. We shall see that extended nonequilibrium does not
relax in general, and is in fact unstable. On this basis it will be argued
that Bohm's dynamics is untenable, as there would be no reason to expect to
see quantum equilibrium in our world today.

In Section 2 we formally introduce the notion of extended nonequilibrium in
Bohm's dynamics. In Section 3 we compare and contrast Bohm's dynamics with
classical dynamics, and for the former we show that there exist \textit{two}
distinct equilibrium distributions in phase space. In Section 4 we compare and
contrast Bohm's dynamics with de Broglie's dynamics for a simple example: a
particle in the ground state of a bound system. This example is unrealistic
and does not by itself enable any significant conclusions to be drawn; but it
serves an illustrative purpose.

In Section 5 we consider more realistic examples of systems with wave
functions that are superpositions of energy eigenstates. We consider the
harmonic oscillator and the hydrogen atom, for specific superpositions, and we
show by numerical simulations that quantum equilibrium is unstable for these
systems. We then consider the oscillator for an arbitrary superposition (with
a bounded energy spectrum), and we provide an analytic proof that the system
is unstable for asymptotically large initial positions. Because the harmonic
oscillator occurs in many key areas of physics -- including field theory -- we
may conclude that in Bohm's dynamics there is no general tendency to relax to
quantum equilibrium and that the quantum equilibrium state is in fact unstable.

In Section 6 we show that a similar instability occurs if one applies Bohm's
dynamics to high-energy field theory in the early universe. We conclude that
if the universe started in a nonequilibrium state, and if it were governed by
Bohm's dynamics, then we would not see equilibrium today. In particular, there
would be no bound atomic states and even the vacuum would contain arbitrarily
large field strengths, in sharp conflict with observation.

Finally, in Section 7 we draw the conclusion that, while de Broglie's dynamics
is a tenable physical theory, Bohm's dynamics is not.

\section{Extended nonequilibrium in Bohm's dynamics}

In phase space the quantum equilibrium (or quantum theoretical) distribution
is%
\begin{equation}
\rho_{\mathrm{QT}}(q,p,t)=\left\vert \Psi(q,t)\right\vert ^{2}\delta
^{3N}(p-\nabla_{q}S(q,t))\ , \label{equ1}%
\end{equation}
where (again) $q=(\mathbf{x}_{1},\mathbf{x}_{2},...,\mathbf{x}_{N})$,
$\nabla_{q}=(\mathbf{\nabla}_{1},\mathbf{\nabla}_{2},...,\mathbf{\nabla}_{N})$
and where $p=(\mathbf{p}_{1},\mathbf{p}_{2},...,\mathbf{p}_{N})$. As we have
seen, according to Bohm's dynamics this distribution will hold at all $t$ if
it holds at $t=0$.

However, in principle Bohm's dynamics allows arbitrary initial distributions
$\rho(q,p,0)$ on phase space whose time evolution $\rho(q,p,t)$ will be given
by the continuity equation%
\begin{equation}
\frac{\partial\rho}{\partial t}+\nabla_{q}\cdot(\rho\dot{q})+\nabla_{p}%
\cdot(\rho\dot{p})=0\ .
\end{equation}
Here $\nabla_{p}$ denotes a $3N$-dimensional gradient with respect to the
momenta. The phase-space velocity field%
\begin{equation}
(\dot{q},\dot{p})=(\mathbf{\dot{x}}_{1},\mathbf{\dot{x}}_{2},...,\mathbf{\dot
{x}}_{N},\mathbf{\dot{p}}_{1},\mathbf{\dot{p}}_{2},...,\mathbf{\dot{p}}_{N})
\end{equation}
has components ($i=1,2,...,N$)%
\begin{equation}
\mathbf{\dot{x}}_{i}=\mathbf{p}_{i}/m_{i},\ \ \ \ \ \mathbf{\dot{p}}%
_{i}=-\mathbf{\nabla}_{i}(V+Q)\ .
\end{equation}

The key question is whether `reasonable' initial nonequilibrium distributions%
\begin{equation}
\rho(q,p,0)\neq\left\vert \Psi(q,0)\right\vert ^{2}\delta^{3N}(p-\nabla
_{q}S(q,0))
\end{equation}
tend to relax to (extended) quantum equilibrium or not. We shall present
strong evidence that they do not.

\section{Comparison with classical dynamics}

Bohm's dynamics is, in effect, just Newton's dynamics with an additional
time-dependent potential $Q(q,t)$ added to the usual classical potential
function $V$. Equivalently, it is a Hamiltonian dynamics with a classical
Hamiltonian%
\begin{equation}
H=\sum_{i=1}^{N}\frac{\mathbf{p}_{i}^{2}}{2m_{i}}+V+Q\ ,
\end{equation}
where $\dot{q}=\nabla_{p}H$ and $\dot{p}=-\nabla_{q}H$. Because of the
explicit time dependence of $Q$, the energy of a system of particles is not
conserved in general. Specifically, if we take $H$ to be the total energy then
$dH/dt=\partial Q/\partial t$ -- which is generally non-zero. Thus the
trajectories are not confined to a fixed energy surface in phase space.

However we still have Liouville's theorem, just as for any Hamiltonian system.
The total time derivative of $\rho(q,p,t)$ is given by%
\begin{align*}
\frac{d\rho}{dt}  &  =(\nabla_{q}\rho)\cdot\dot{q}+(\nabla_{p}\rho)\cdot
\dot{p}+\frac{\partial\rho}{\partial t}\\
&  =(\nabla_{q}\rho)\cdot\dot{q}+(\nabla_{p}\rho)\cdot\dot{p}-\nabla_{q}%
\cdot(\rho\dot{q})-\nabla_{p}\cdot(\rho\dot{p})
\end{align*}
and so along a trajectory we have%
\begin{equation}
\frac{d\rho}{dt}=-\rho\left(  \nabla_{q}\cdot\dot{q}+\nabla_{p}\cdot\dot
{p}\right)  =0\ . \label{Liou}%
\end{equation}

This implies that the dynamics contains \textit{two }equilibrium
distributions. For if $\rho(q,p,0)=c$ (for some constant $c$) over the
available region of phase space, then (\ref{Liou}) implies that%
\begin{equation}
\rho(q,p,t)=c \label{equ2}%
\end{equation}
at all times $t$. This is just the usual (classical) equilibrium distribution.
On the other hand, the quantum equilibrium distribution (\ref{equ1}) is also
conserved by Bohm's dynamics.

The existence of two distinct equilibrium states raises the question: will
there be a tendency for relaxation to occur to one of these equilibrium
states, or to neither? One might guess that the existence of two equilibrium
measures will in some sense `confuse' the system.

It might also be suggested that, because the support of the quantum
equilibrium distribution (\ref{equ1}) has zero Lebesgue measure in phase
space, and because phase-space volume is conserved by a Hamiltonian flow, then
an initial distribution with finite Lebesgue measure will not be able to relax
to quantum equilibrium. However, the conservation of phase-space volume does
not by itself rule out the possibility that an initial distribution with
finite Lebesgue measure could approach quantum equilibrium in the
infinite-time limit, by an appropriate `squeezing' of the evolving
distribution with respect to dimensions that are orthogonal to the surface
$p=\nabla_{q}S$ in phase space, together with a simultaneous unlimited
spreading of the distribution over the whole of that surface (which can be
infinitely extended) in such a way as to conserve the total phase-space
volume. (We shall see that such `squeezing' does occur to some degree in some
circumstances -- see Figure 2 -- but does not appear to be generic.)

Thus, given our experience with classical systems, it is not immediately
obvious how the above system will behave. Bohm's dynamics defines an unusual
dynamical system, and one ought to beware of standard intuitions and expectations.

\section{Comparison with de Broglie's dynamics for the ground state of a bound
system}

As a simple and preliminary example, consider a single particle in the ground
state of a bound system -- such as a hydrogen atom or a simple harmonic
oscillator. This example is unrealistic but serves an illustrative purpose.
The wave function may be written in the form%
\[
\psi(\mathbf{x},t)=\phi_{0}(\mathbf{x})e^{-iE_{0}t}\ ,
\]
where $\phi_{0}(\mathbf{x})$ is a real and non-negative eigenfunction of the
Hamiltonian operator%
\[
\hat{H}=-\frac{1}{2m}\nabla^{2}+V
\]
and $E_{0}$ is the ground-state energy eigenvalue. We take $\phi
_{0}(\mathbf{x})$ to be localised around the origin at $\mathbf{x}=0$.

Because $\phi_{0}$ is real the phase gradient vanishes, $\mathbf{\nabla}S=0$.
Furthermore, because $\phi_{0}$ is non-negative we have $|\psi|=\phi_{0}$ and
so the eigenvalue equation $\hat{H}\phi_{0}=E_{0}\phi_{0}$ implies that%
\[
V+Q=E_{0}%
\]
(for all $\mathbf{x}$).

Now, let us first consider the behaviour of this system according to de
Broglie's dynamics. We have $\mathbf{p}=\mathbf{\nabla}S=0$ everywhere, so
that the velocity of the bound particle vanishes no matter where it happens to
be located. If we then consider an initial ensemble of such particles, whose
positions have an initial distribution $\rho_{0}(\mathbf{x})$ that deviates
\textit{slightly} from the equilibrium distribution $|\phi_{0}(\mathbf{x}%
)|^{2}$, then because the particles are at rest we have $\rho(\mathbf{x}%
,t)=\rho_{0}(\mathbf{x})$ for all $t$ and we deduce (trivially) that initial
small deviations from equilibrium will remain small (and indeed static). On
the other hand, for initial wave functions that are superpositions of
different energy eigenfunctions, extensive numerical evidence shows that
initial small deviations from equilibrium quickly relax, with $\rho
(\mathbf{x},t)$ rapidly approaching $|\psi(\mathbf{x},t)|^{2}$ (on a
coarse-grained level, assuming that the initial state has no fine-grained
micro-structure) (Valentini 1992, 2001; Valentini and Westman 2005;
Efthymiopoulos and Contopoulos 2006; Towler, Russell and Valentini 2012; Colin 2012).

In the case of Bohm's dynamics, in contrast, we have%
\[
-\mathbf{\nabla}(V+Q)=-\mathbf{\nabla}E_{0}=0
\]
everywhere, so that now the \textit{acceleration} vanishes no matter where the
particle happens to be located. Therefore, an initial small deviation of the
momentum $\mathbf{p}$ from $\mathbf{\nabla}S$ -- that is, an initial small
deviation of $\mathbf{p}$ from $0$ -- remains small (and indeed static).
However, a small (and constant) non-zero value of $\mathbf{p}$ will cause an
unbounded growth in nonequilibrium with respect to \textit{position}. For
example, let the initial position distribution $\rho_{0}(\mathbf{x})$ be
concentrated in a small region around some point $\mathbf{x}=\mathbf{x}_{0}$
close to the origin, and assume that each particle in the ensemble has the
same non-zero value of $\mathbf{p}$ pointing away from the origin. Then each
particle will move away from the origin at a uniform speed $|\mathbf{p}|/m$
and the distribution at time $t$ will be simply $\rho(\mathbf{x},t)=\rho
_{0}(\mathbf{x}-(\mathbf{p}/m)t)$ -- that is, at time $t$ the distribution
will be concentrated in a small region around the point $\mathbf{x}%
=\mathbf{x}_{0}+(\mathbf{p}/m)t$, which moves at uniform speed away from the
origin, implying an ever larger deviation of $\rho$ from equilibrium. Thus,
for this simple case, the bound state becomes unbound and the quantum
equilibrium state is unstable.

Similarly, one may also consider excited energy eigenstates. In de Broglie's
dynamics, the trajectories for such states are generally too simple for
relaxation to occur. In Bohm's dynamics, it may be shown that the bound state
again becomes unbound when the initial particle momentum is sufficiently large
(Goldstein and Struyve 2014).

It should be emphasised that these features of the ground state (and of
excited states) do not by themselves present a difficulty, neither for de
Broglie's dynamics nor for Bohm's, because it is completely unrealistic for a
system in nature to occupy an energy eigenstate for an indefinite period of
time. All physical systems that we have access to have a long and violent
astrophysical history that ultimately traces back to the big bang. A hydrogen
atom, for example, will have undergone interactions in its past and its wave
function will have been a \textit{superposition} of many energy eigenstates.
Even if the atom is presently in an energy eigenstate, in the past it will not
have been. Thus the above features of energy eigenstates are not relevant to
the empirical adequacy of either version of the dynamics.

In the case of de Broglie's dynamics, we know that in the past when an atom
was in a state of superposition it will have undergone rapid relaxation to
quantum equilibrium. Therefore, the fact that relaxation does not occur for
the ground state (or for excited states) is in no way a difficulty for de
Broglie's dynamics. At first sight, then, it may seem entirely possible that
the same could be true for Bohm's dynamics: quantum equilibrium is unstable
for the ground state (and for excited states) but it might not be for the more
realistic case of superpositions. However, as we shall now show, for Bohm's
dynamics similar results are obtained for superpositions. Thus, even for
realistic initial states there is no relaxation in Bohm's dynamics. Bound
states become unbound, and the quantum equilibrium state is unstable.

\section{Instability of Bohm's dynamics for superpositions of energy
eigenstates}

We shall first demonstrate, by means of numerical simulations for two
examples, that in Bohm's dynamics quantum equilibrium is unstable for wave
functions that are superpositions of energy eigenstates. In particular, we
consider the harmonic oscillator and the hydrogen atom for specific
superpositions. We then consider arbitrary superpositions for the oscillator,
with a bounded energy spectrum, and we show analytically that the system is
unstable in the asymptotic limit of large initial positions.

\subsection{Numerical results for the harmonic oscillator}

In our first example we consider a one-dimensional harmonic oscillator with an
initial wave function that is a superposition of the first three energy
eigenstates. The superposition is equally weighted, with randomly-chosen
initial phases. We take units such that $\hbar=m=\omega_{0}=1$, where
$\omega_{0}$ is the angular frequency. The classical potential is then
$V=\frac{1}{2}x^{2}$.

The Schr\"{o}dinger equation for the system reads%
\[
i\frac{\partial\psi}{\partial t}=-\frac{1}{2}\frac{\partial^{2}\psi}{\partial
x^{2}}+\frac{1}{2}x^{2}\psi\ .
\]
We consider an example with a wave function%
\begin{equation}
\psi(x,t)=\frac{1}{\sqrt{3}}\left(  \phi_{0}e^{-it/2}+e^{i\theta_{1}}\phi
_{1}e^{-i3t/2}+e^{i\theta_{2}}\phi_{2}e^{-i5t/2}\right)  \ , \label{psi}%
\end{equation}
where $\phi_{0}$, $\phi_{1}$, $\phi_{2}$ are the first three energy
eigenstates of the oscillator and $\theta_{1}$, $\theta_{2}$ are
randomly-chosen initial phases. According to Bohm's dynamics the acceleration
of the particle is given by%
\begin{equation}
a\equiv\ddot{x}=-x-\frac{\partial Q}{\partial x}\ , \label{accSHO}%
\end{equation}
where here the quantum potential $Q\equiv-(1/2)(1/|\psi|)\partial^{2}%
|\psi|/\partial x^{2}$ is periodic in time with period $2\pi$.

Given the expression (\ref{psi}) for $\psi$, we may plot the acceleration
field $a=a(x,t)$ numerically. We find that in the region $x>3$ the
acceleration satisfies\footnote{We have verified this numerically up to
$x=10^{3}$. An analytical proof of similar behaviour, for an arbitrary
superposition and for asymptotically large $x$, is given below.}%
\begin{equation}
a>-\frac{2}{x^{2}}\ . \label{lba}%
\end{equation}
This is shown in Figure 1, where we plot $a+2/x^{2}$ for $t$ in $(0,2\pi)$ and
$x$ in $(3,10)$ (taking $\theta_{1}=1.1$, $\theta_{2}=1.8$).%

\begin{figure}
\begin{center}
\includegraphics[width=\textwidth]%
{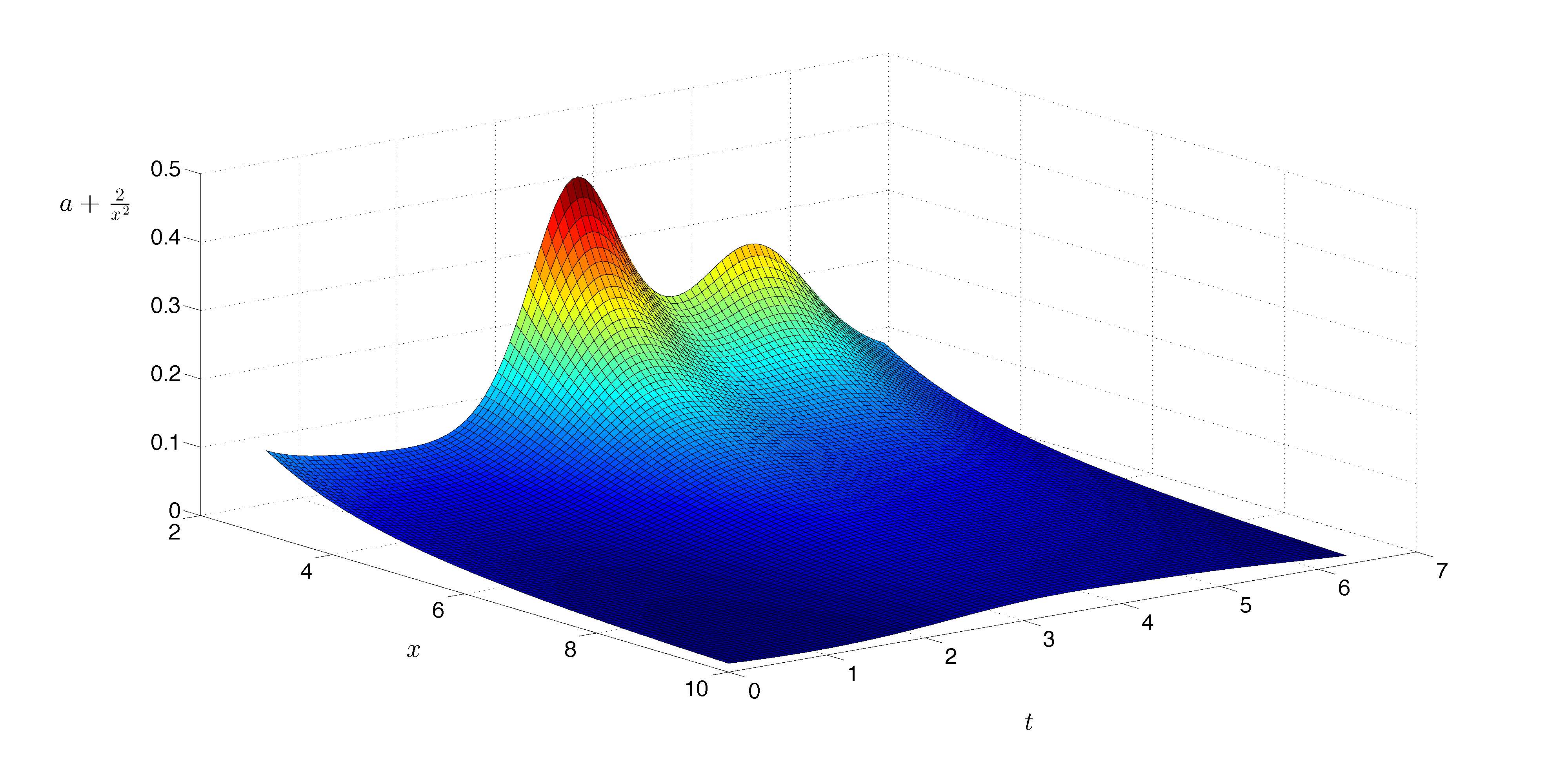}%
\caption{Plot of $a+2/x^{2}$ for $t$ in $(0,2\pi)$ and $x$ in $(3,10)$,
showing that $a+2/x^{2}>0$ in this region.}%
\end{center}
\end{figure}

Now it is an elementary property of Newtonian dynamics that if $\ddot
{x}=-b/x^{2}$ (for some constant $b>0$), and if the particle begins at an
initial point $x_{0}>0$ with an initial velocity $v_{0}$ greater than the
`escape velocity' $v_{\mathrm{escape}}=\sqrt{2b/x_{0}}$, then the particle
will escape to infinity: $x(t)\rightarrow\infty$ as $t\rightarrow\infty$.
Clearly the same conclusion will hold if $\ddot{x}=-b/x^{2}+\xi$ where $\xi
>0$, since $\xi$ amounts to an additional force directed away from the origin.

Thus in the example given, if the particle begins in the region $x>3$ with an
initial momentum $p_{0}>p_{\mathrm{escape}}=\sqrt{4/x_{0}}$ (taking $b=2$),
then it will escape to infinity. While the region $x>3$ does not include the
bulk of the support of the initial packet (located around the origin with a
spread of order $\sim1$), it is not so far out in the tail as to be negligible.

We may conclude that, for this example, quantum equilibrium is unstable under
Bohm's dynamics. To illustrate this more explicitly, we may calculate the time
evolution of some particular initial distributions in phase space (now taking
initial phases $\theta_{1}=2$, $\theta_{2}=4$). In Figure 2 we show an initial
distribution at $t=0$ that is concentrated in a small region of phase space
centred on a point of the curve $p=\partial S(x,0)/\partial x$. After a time
$t=5$, we find that the evolved distribution is bunched around the curve
$p=\partial S(x,5)/\partial x$. A significant relaxation towards quantum
equilibrium has clearly occurred. In contrast, in Figure 3 we show an initial
distribution that is the same as before but displaced along the $p$-axis by
$+0.5$. Again calculating up to a time $t=5$, we now find that the
distribution has departed further from the curve $p=\partial S(x,5)/\partial
x$ and the particles appear to be escaping. These simulations illustrate how
the particles will escape if their initial momenta are sufficiently large.%

\begin{figure}
\begin{center}
\includegraphics[width=\textwidth]%
{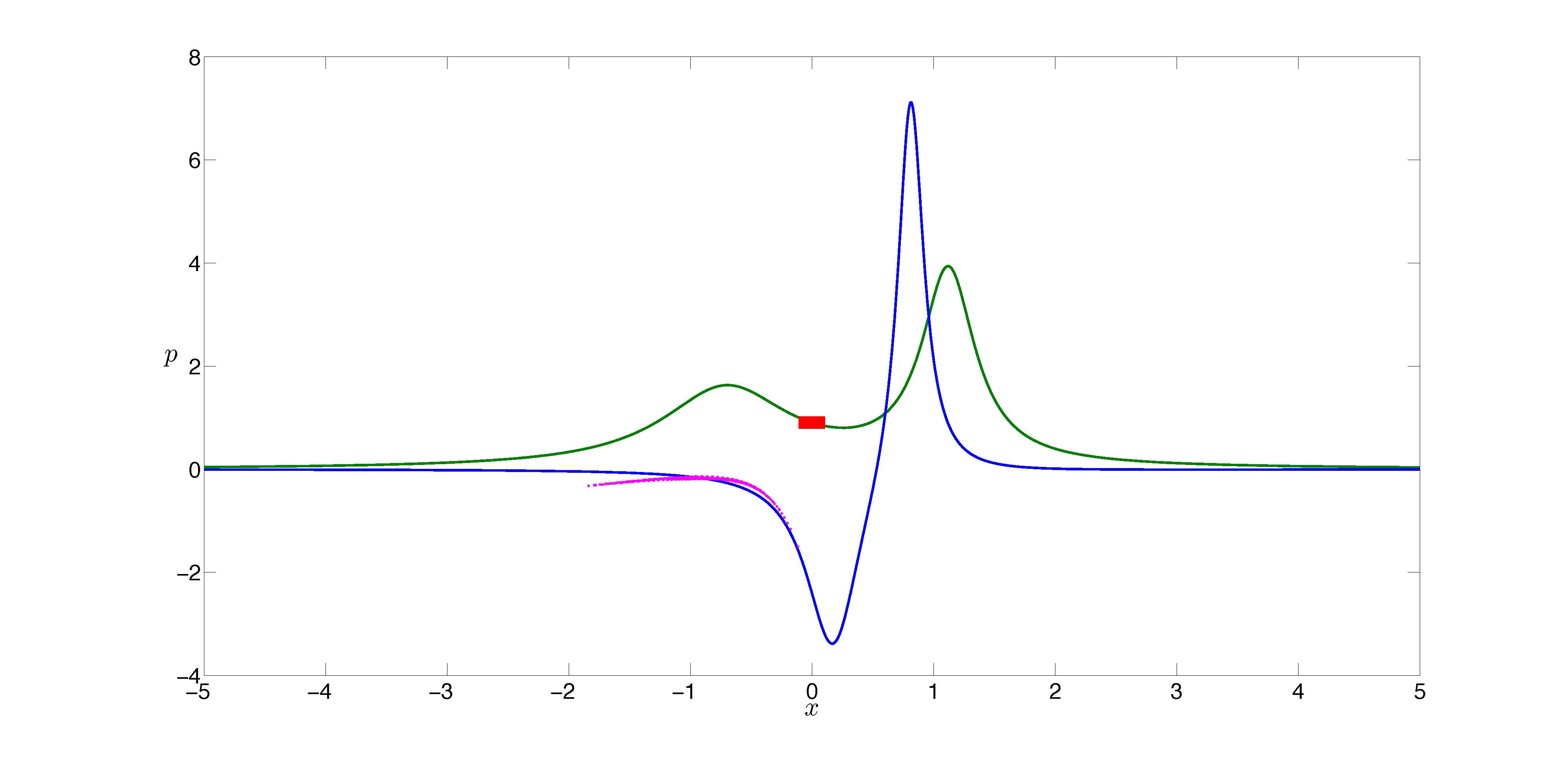}%
\caption{An initial distribution (red) at $t=0$ that is concentrated in a
small region of phase space centred on a point of the curve $p=\partial
S(x,0)/\partial x$ (green). At $t=5$ the evolved distribution (magenta) is
bunched around the curve $p=\partial S(x,5)/\partial x$ (blue).}%
\end{center}
\end{figure}
%

\begin{figure}
\begin{center}
\includegraphics[
width=\textwidth]%
{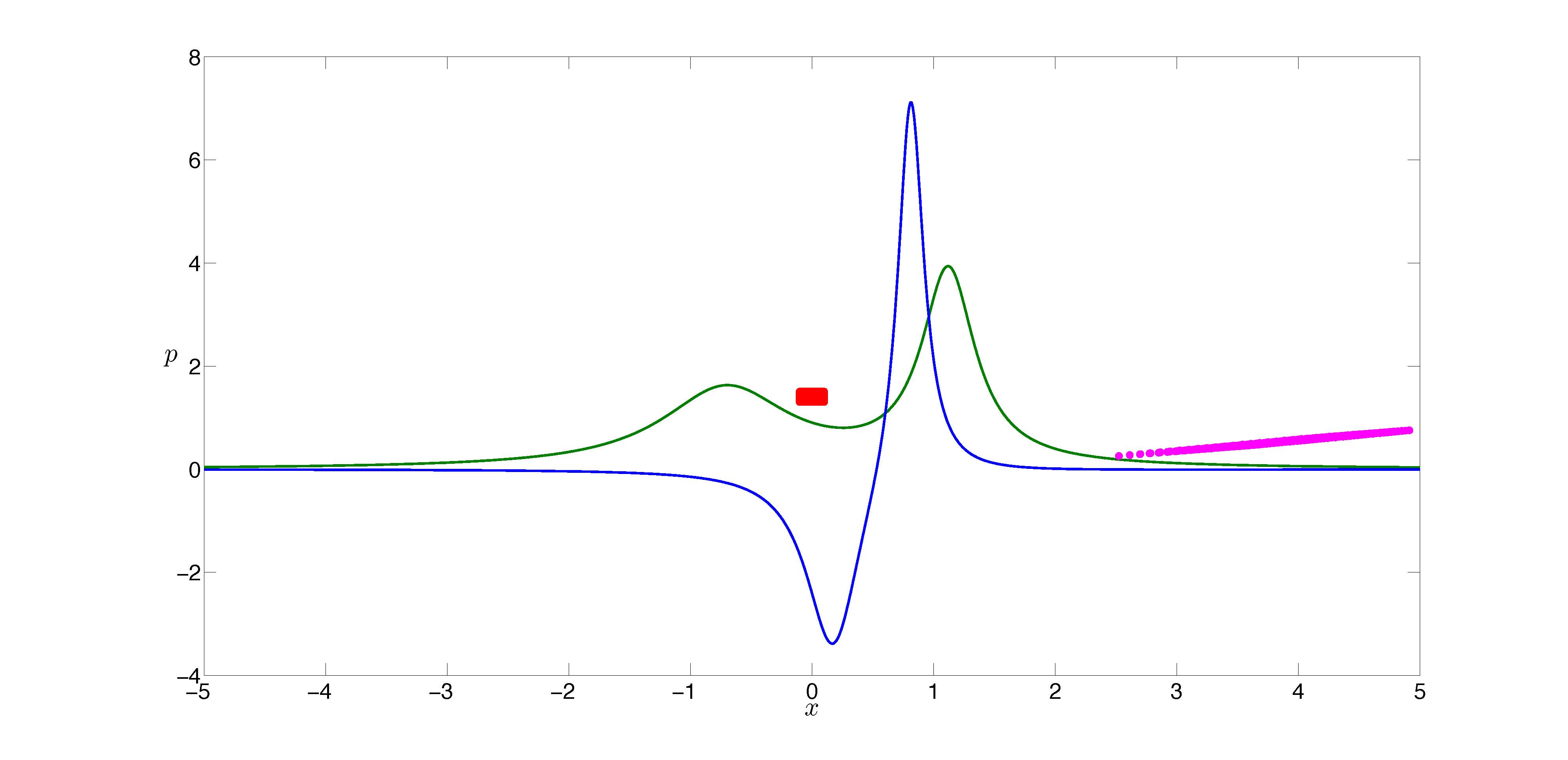}%
\caption{An initial distribution (red) at $t=0$ that is the same as in Figure
2 but displaced along the $p$-axis by $+0.5$. At $t=5$ the distribution
(magenta) has departed further from the curve $p=\partial S(x,5)/\partial x$
(blue).}%
\end{center}
\end{figure}

\subsection{Numerical results for the hydrogen atom}

In our second example we consider a hydrogen-like atom -- a (spinless)
particle moving in three dimensions in a Coulomb potential. The initial wave
function $\psi(\mathbf{x},0)$ is chosen to be a superposition%
\[
\psi(\mathbf{x},0)=\frac{1}{\sqrt{3}}\left[  \phi_{100}(\mathbf{x})+e^{i}%
\phi_{211}(\mathbf{x})+e^{2i}\phi_{32-1}(\mathbf{x})\right]
\]
of three energy eigenstates $\phi_{nlm}(\mathbf{x})$. We have calculated some
of the particle trajectories numerically according to Bohm's dynamics. We now
take units such that $\hbar=m=a_{0}=1$, where $a_{0}$ is the Bohr radius.

A representative sample of our results is displayed in Figure 4. In all the
cases shown the initial position is taken to be $\mathbf{x}_{0}=(x_{0}%
,y_{0},z_{0})=(0.5,0.5,0.5)$. Five trajectories are plotted. For comparison,
the trajectory in black is obtained from de Broglie's dynamics (by integrating
$m\mathbf{\dot{x}}=\mathbf{\nabla}S$).\footnote{We are grateful to Ward
Struyve for independently checking the accuracy of this trajectory.} The other
four trajectories are obtained from Bohm's dynamics (by integrating
$\mathbf{\ddot{x}}=-\mathbf{\nabla}(V+Q)$), with varying values of initial
momentum. The trajectory in blue has an initial momentum equal to the de
Broglie value, $\mathbf{p}_{0}^{\mathrm{deB}}=\mathbf{\nabla}S_{0}%
(\mathbf{x}_{0})=(-0.19,-0.11,-0.02)$. The blue and black trajectories are
identical, as they must be. The trajectory in green has an initial momentum
$\mathbf{p}_{0}=(-0.2,-0.1,0)$ (or $\mathbf{p}_{0}=\mathbf{p}_{0}%
^{\mathrm{deB}}+(-0.01,0.01,0.02)$), which differs only slightly from
$\mathbf{p}_{0}^{\mathrm{deB}}$. The trajectory in magenta has an initial
momentum $\mathbf{p}_{0}=\mathbf{p}_{0}^{\mathrm{deB}}+(0.05,0.05,0.05)$,
while the trajectory in red has an initial momentum $\mathbf{p}_{0}%
=\mathbf{p}_{0}^{\mathrm{deB}}+(0.1,0.1,0.1)$.%

\begin{figure}
\begin{center}
\includegraphics[
width=\textwidth
]%
{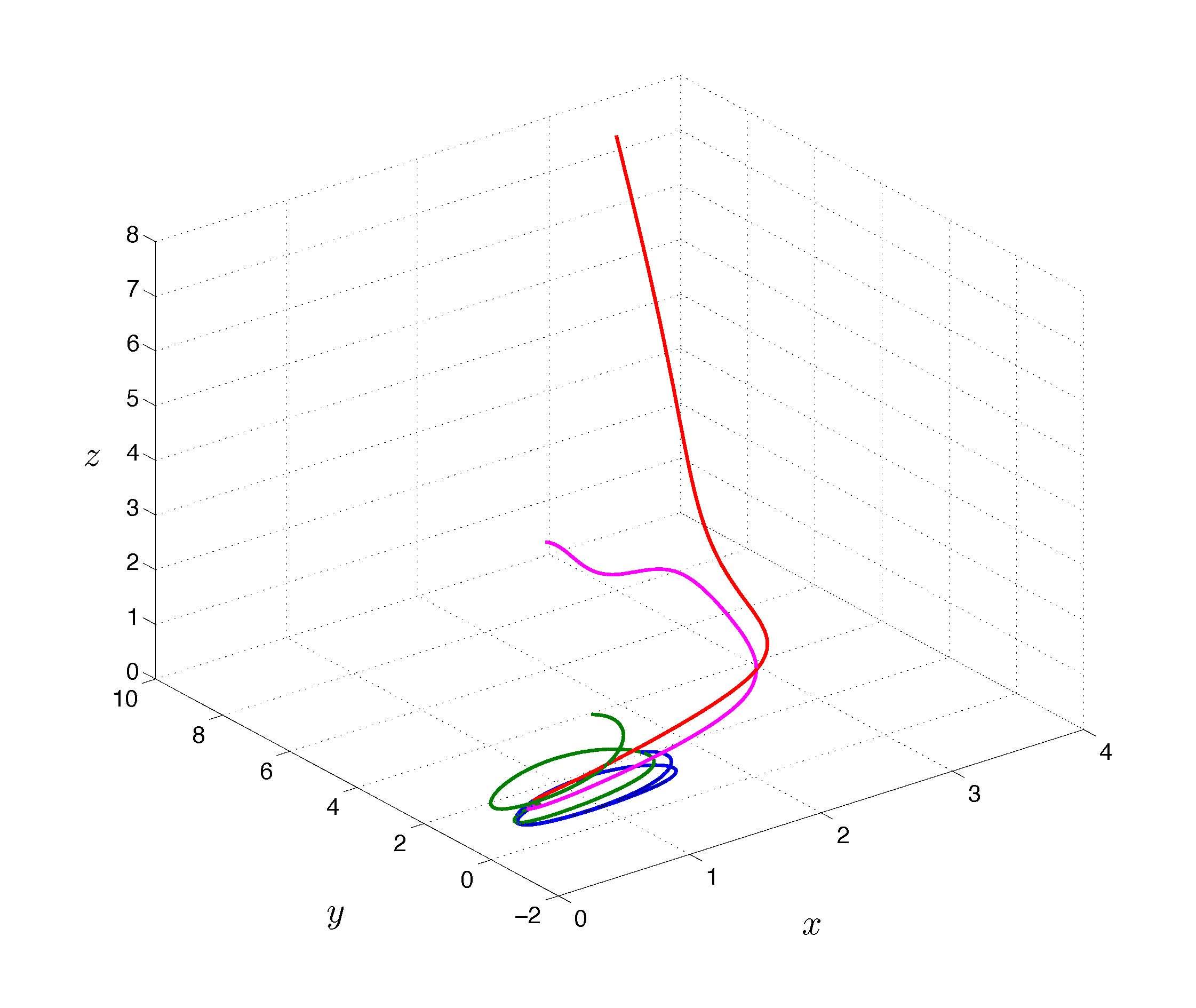}%
\caption{Trajectories for the hydrogen atom.}%
\end{center}
\end{figure}

The results speak for themselves. Perturbing the initial momentum only
slightly away from the de Broglie value yields a noticeable but fairly small
change in the trajectory (in green), with the particle appearing to remain
bound. Adding a somewhat larger perturbation makes the orbit (in magenta)
extend far away from the bulk of the wave packet, while a still larger
perturbation results in a trajectory (in red) that appears to leave the system altogether.

Of course, strictly speaking, because we have not integrated all the way to
$t=\infty$ these results are not a completely rigorous proof that some
trajectories (with initial momenta differing substantially from the de Broglie
value) actually escape to infinity. Such a proof will now be given -- for
arbitrary states of the oscillator.

\subsection{Asymptotic instability for arbitrary states of the harmonic
oscillator}

So far we have demonstrated the instability of Bohm's dynamics numerically,
and for certain specific superpositions, for both the harmonic oscillator and
the hydrogen atom. Here we provide an analytical proof of instability for the
oscillator -- for arbitrary states with a bounded energy spectrum, and in the
asymptotic limit of large initial positions $x_{0}$.

Consider again the (one-dimensional) oscillator, but now with an arbitrary
superposition%
\begin{equation}
\psi(x,t)=\sum_{m=0}^{M}c_{m}(0)e^{-i(m+1/2)t}\phi_{m}(x) \label{supn}%
\end{equation}
of energy eigenstates%
\begin{equation}
\phi_{m}(x)=\frac{1}{\pi^{1/4}}\frac{1}{\sqrt{2^{m}m!}}\mathcal{H}%
_{m}(x)e^{-x^{2}/2}\ ,
\end{equation}
up to some maximal eigenvalue $E_{M}$, where $\mathcal{H}_{m}(x)$ is the
Hermite polynomial of order $m$. It will be shown that for large and positive
$x$ the acceleration field \textit{always} has an asymptotic lower bound%
\begin{equation}
a\gtrsim-\frac{b}{x^{2}}\ , \label{alb}%
\end{equation}
where%
\begin{equation}
b=\frac{\left\vert c_{M-1}(0)\right\vert }{\left\vert c_{M}(0)\right\vert
}\sqrt{\frac{M}{2}} \label{b}%
\end{equation}
is a positive constant that is determined by the superposition (\ref{supn}).

To derive the asymptotic lower bound let us write%
\[
\left\vert \psi(x,t)\right\vert ^{2}=e^{-x^{2}}P(x,t)\ ,
\]
where%
\[
P(x,t)=\sum_{n=0}^{N}\alpha_{n}(t)x^{n}%
\]
is a positive polynomial of order $N=2M$. The acceleration field
(\ref{accSHO}) may be written purely in terms of $P$ and its spatial
derivatives:%
\begin{equation}
a=\frac{P^{2}P^{\prime\prime\prime}+(P^{\prime})^{3}-2PP^{\prime}%
P^{\prime\prime}-2xP^{2}P^{\prime\prime}-2P^{2}P^{\prime}+2xP(P^{\prime})^{2}%
}{4P^{3}}\ . \label{aP}%
\end{equation}
(To show this it is useful to write $Q\equiv-(1/2\left\vert \psi\right\vert
)\partial^{2}\left\vert \psi\right\vert /\partial x^{2}$ in the form
$Q=-(1/4|\psi|^{2})(\partial^{2}|\psi|^{2}/\partial x^{2})+(1/8|\psi
|^{4})(\partial|\psi|^{2}/\partial x)^{2}$.)

The denominator in (\ref{aP}) will certainly contain the term $4\alpha_{N}%
^{3}x^{3N}$ (assuming $\alpha_{N}\neq0$). The numerator can contain at most a
term proportional to $x^{3N-1}$, coming from the last three terms in the
numerator. However the coefficient will be given by%
\[
-2\alpha_{N}^{3}N(N-1)-2\alpha_{N}^{3}N+2\alpha_{N}^{3}N^{2}=0\ ,
\]
so the leading term in the numerator will in fact be proportional to (at most)
$x^{3N-2}$. This term also comes from the last three terms in the numerator,
and its coefficient is%
\begin{equation}
-2\alpha_{N}^{2}\alpha_{N-1}\ . \label{coeff}%
\end{equation}
Thus, assuming that the coefficient (\ref{coeff}) is non-zero, for large $x$
we will have the asymptotic behaviour%
\begin{equation}
a\sim-\frac{\alpha_{N-1}}{2\alpha_{N}}\frac{1}{x^{2}}\ . \label{asymp}%
\end{equation}
(If instead (\ref{coeff}) vanishes, we will have $a\sim c(t)/x^{p}$ with
$p\geq3$ and where $c(t)$ is some bounded function of time.)

Now the coefficient $\alpha_{N}$ is given by%
\[
\alpha_{N}=\left\vert c_{M}(0)\right\vert ^{2}\frac{1}{\sqrt{\pi}}\frac
{1}{2^{M}M!}\left[  \mathrm{coeff}(\mathcal{H}_{M}(x),M)\right]  ^{2}\ ,
\]
where $\mathrm{coeff}(P(x),k)$ is the coefficient of the term of order $k$ in
the polynomial $P(x)$. As for the coefficient $\alpha_{N-1}$, the term
containing it can only come from a product of $\mathcal{H}_{M}$ with
$\mathcal{H}_{M-1}$. (The product of $\mathcal{H}_{M}$ with itself will
produce no such term because $\mathcal{H}_{M}$ contains no term of order
$M-1$.) Writing $c_{m}(0)=\left\vert c_{m}(0)\right\vert e^{i\theta_{m}}$, we
find that%
\begin{align*}
\alpha_{N-1}  &  =2\left\vert c_{M}(0)\right\vert \left\vert c_{M-1}%
(0)\right\vert \sqrt{\frac{2M}{\pi}}\frac{1}{2^{M}M!}\mathrm{coeff}%
(\mathcal{H}_{M}(x),M)\\
&  \times\mathrm{coeff}(\mathcal{H}_{M-1}(x),M-1)\cos\left(  t+(\theta
_{M}-\theta_{M-1})\right)  \ .
\end{align*}

Since $\mathrm{coeff}(\mathcal{H}_{M-1}(x),M-1)=\frac{1}{2}\mathrm{coeff}%
(\mathcal{H}_{M}(x),M)$, the asymptotic behaviour (\ref{asymp}) of the
acceleration field is then found to be%
\begin{equation}
a\sim-\frac{\left\vert c_{M-1}(0)\right\vert }{\left\vert c_{M}(0)\right\vert
}\sqrt{\frac{M}{2}}\frac{1}{x^{2}}\cos\left(  t+(\theta_{M}-\theta
_{M-1})\right)  \ .
\end{equation}
Because the cosine oscillates between $+1$ and $-1$, we indeed have the
asymptotic lower bound (\ref{alb}) with the coefficient (\ref{b}).

Thus for sufficiently large initial positions $x_{0}$ we may again deduce that
if the initial velocity $v_{0}$ is larger than $v_{\mathrm{escape}}%
=\sqrt{2b/x_{0}}$ -- with $b$ now given by (\ref{b}) -- the particle will
escape to infinity. (Note that $v_{\mathrm{escape}}\rightarrow0$ as
$x_{0}\rightarrow\infty$.)

For an initial quantum equilibrium ensemble, there will always be some small
but finite fraction of points that begin far out in the tail in position
space. If the initial velocities of such points are slightly perturbed away
from the equilibrium de Broglie values, in such a way that they exceed the
small threshold $v_{\mathrm{escape}}$, then that fraction of the ensemble will escape.

This asymptotic result has been proved analytically. The numerical simulations
of Section 5.1 demonstrate a similar behaviour in a region close to the bulk
of the initial packet. Taken together, these results constitute strong
evidence that the instability of quantum equilibrium is generic for the
harmonic oscillator as described by Bohm's dynamics. We may expect that such
instability will occur for this system with any physically-reasonable initial
wave function. This is an important conclusion because the harmonic oscillator
is a fundamental system that occurs in many key areas of physics -- including
field theory.

\section{Cosmology and field theory}

The above results provide strong evidence that there is no tendency to relax
to quantum equilibrium in Bohm's dynamics, and that the quantum equilibrium
state is in fact unstable. It is then reasonable to conclude that if the
universe started in a nonequilibrium state -- and if the universe were
governed by Bohm's dynamics -- then we would \textit{not} see quantum
equilibrium today. The Born rule for particle positions would fail, momenta
would take non-quantum-mechanical values, and there would be no bound states
such as atoms or nuclei.

As a counter-argument, it might be suggested that the early universe could
reach equilibrium long before atoms form (about 400,000 years after the big
bang, when electrons and protons combine to form neutral hydrogen). Stable
bound states would then form at later times in the usual way. Furthermore,
since our discussion so far has been confined to the low-energy domain, it
might be thought that conclusions about the early (very hot) universe are in
any case unwarranted. Both objections may be overcome by showing that the same
instability appears if one applies Bohm's dynamics to high-energy field theory.

In classical field theory a free, minimally-coupled, massless scalar field
$\phi$, on a curved spacetime with 4-metric $g^{\mu\nu}$, has a Lagrangian
density $\mathcal{L}=\frac{1}{2}\sqrt{-g}g^{\mu\nu}\partial_{\mu}\phi
\partial_{\nu}\phi$ (where $\mu,\nu=0,1,2,3$). On an expanding flat space with
line element%
\[
d\tau^{2}=dt^{2}-a^{2}d\mathbf{x}^{2}%
\]
(where $a=a(t)$ is the scale factor), we have%
\[
\mathcal{L}=\tfrac{1}{2}a^{3}\dot{\phi}^{2}-\tfrac{1}{2}a(\mathbf{\nabla}%
\phi)^{2}\ .
\]
It is convenient to work in Fourier space, in terms of Fourier components%
\[
\phi_{\mathbf{k}}(t)=\frac{1}{(2\pi)^{3/2}}\int d^{3}\mathbf{x}\;\phi
(\mathbf{x},t)e^{-i\mathbf{k}\cdot\mathbf{x}}\ .
\]
These may be written as%
\[
\phi_{\mathbf{k}}=\frac{\sqrt{V}}{(2\pi)^{3/2}}\left(  q_{\mathbf{k}%
1}+iq_{\mathbf{k}2}\right)
\]
for real $q_{\mathbf{k}r}$ ($r=1$, $2$), where $V$ is a box normalisation
volume. The Lagrangian $L=\int d^{3}\mathbf{x}\;\mathcal{L}$ then becomes%
\[
L=\sum_{\mathbf{k}r}\frac{1}{2}\left(  a^{3}\dot{q}_{\mathbf{k}r}^{2}%
-ak^{2}q_{\mathbf{k}r}^{2}\right)  \ .
\]
Introducing the canonical momenta $\pi_{\mathbf{k}r}\equiv\partial
L/\partial\dot{q}_{\mathbf{k}r}=a^{3}\dot{q}_{\mathbf{k}r}$, the Hamiltonian
becomes%
\[
H=\sum_{\mathbf{k}r}\left(  \frac{1}{2a^{3}}\pi_{\mathbf{k}r}^{2}+\frac{1}%
{2}ak^{2}q_{\mathbf{k}r}^{2}\right)  \ .
\]
(Note that, at time $t$, a coordinate distance $|d\mathbf{x}|$ corresponds to
a physical distance $a(t)|d\mathbf{x}|$. It is usual to take $a_{0}=1$ today
(at time $t_{0}$), so that $|d\mathbf{x}|$ is a physical or proper distance
today. Physical wavelengths are given by $\lambda_{\mathrm{phys}}=a(t)\lambda
$, where $\lambda=2\pi/k$ is a proper wavelength today and $k=|\mathbf{k}|$ is
the corresponding wave number. The Hubble parameter $H\equiv\dot{a}/a$ -- not
to be confused with the classical Hamiltonian $H$ -- defines a characteristic
length scale $H^{-1}$, usually called the Hubble radius.)

This system is readily quantised. The Schr\"{o}dinger equation for $\Psi
=\Psi\lbrack q_{\mathbf{k}r},t]$ reads%
\begin{equation}
i\frac{\partial\Psi}{\partial t}=\sum_{\mathbf{k}r}\left(  -\frac{1}{2a^{3}%
}\frac{\partial^{2}}{\partial q_{\mathbf{k}r}^{2}}+\frac{1}{2}ak^{2}%
q_{\mathbf{k}r}^{2}\right)  \Psi\ . \label{Sch2'}%
\end{equation}
This implies the continuity equation%
\[
\frac{\partial\left\vert \Psi\right\vert ^{2}}{\partial t}+\sum_{\mathbf{k}%
r}\frac{\partial}{\partial q_{\mathbf{k}r}}\left(  \left\vert \Psi\right\vert
^{2}\frac{1}{a^{3}}\frac{\partial S}{\partial q_{\mathbf{k}r}}\right)  =0\ ,
\]
from which we may identify the de Broglie velocities%
\begin{equation}
\frac{dq_{\mathbf{k}r}}{dt}=\frac{1}{a^{3}}\frac{\partial S}{\partial
q_{\mathbf{k}r}} \label{deB2}%
\end{equation}
(with $\Psi=\left\vert \Psi\right\vert e^{iS}$). This pilot-wave model has
been applied in a cosmological context by Valentini (2010). A preferred
foliation of spacetime, with time function $t$, has been assumed.

Let us now consider the case of a decoupled mode $\mathbf{k}$. Writing
$\Psi=\psi_{\mathbf{k}}(q_{\mathbf{k}1},q_{\mathbf{k}2},t)\varkappa$, where
$\varkappa$ depends only on degrees of freedom for modes $\mathbf{k}^{\prime
}\neq\mathbf{k}$, it follows from (\ref{Sch2'}), (\ref{deB2}) that
$\psi_{\mathbf{k}}$ satisfies%
\begin{equation}
i\frac{\partial\psi_{\mathbf{k}}}{\partial t}=-\frac{1}{2a^{3}}\left(
\frac{\partial^{2}}{\partial q_{\mathbf{k}1}^{2}}+\frac{\partial^{2}}{\partial
q_{\mathbf{k}2}^{2}}\right)  \psi_{\mathbf{k}}+\frac{1}{2}ak^{2}\left(
q_{\mathbf{k}1}^{2}+q_{\mathbf{k}2}^{2}\right)  \psi_{\mathbf{k}}\ ,
\label{S2D}%
\end{equation}
while the de Broglie velocities for $(q_{\mathbf{k}1},q_{\mathbf{k}2})$ are%
\begin{equation}
\dot{q}_{\mathbf{k}1}=\frac{1}{a^{3}}\frac{\partial s_{\mathbf{k}}}{\partial
q_{\mathbf{k}1}},\ \ \ \ \dot{q}_{\mathbf{k}2}=\frac{1}{a^{3}}\frac{\partial
s_{\mathbf{k}}}{\partial q_{\mathbf{k}2}} \label{deB2D}%
\end{equation}
(with $\psi_{\mathbf{k}}=\left\vert \psi_{\mathbf{k}}\right\vert
e^{is_{\mathbf{k}}}$). These equations are formally the same as those of
pilot-wave dynamics for a nonrelativistic particle with a time-dependent
`mass' $m=a^{3}$ and moving (in the $q_{\mathbf{k}1}-q_{\mathbf{k}2}$ plane)
in a harmonic oscillator potential with time-dependent angular frequency
$\omega=k/a$ (Valentini 2007).

In the short-wavelength limit, $\lambda_{\mathrm{phys}}<<H^{-1}$, we recover
the equations for a decoupled mode $\mathbf{k}$ on Minkowski spacetime --
since, roughly speaking, the timescale $\Delta t\propto\lambda_{\mathrm{phys}%
}$ over which $\psi_{\mathbf{k}}=\psi_{\mathbf{k}}(q_{\mathbf{k}%
1},q_{\mathbf{k}2},t)$ evolves will be much smaller than the expansion
timescale $H^{-1}\equiv a/\dot{a}$ (Valentini 2007, 2008b). On such timescales
$a$ is approximately constant and the equations (\ref{S2D}), (\ref{deB2D})
reduce to those of pilot-wave dynamics for a nonrelativistic particle of
constant mass $m=a^{3}$ moving in an oscillator potential of constant angular
frequency $\omega=k/a$.

We may now readily write down Bohm's dynamics for the same decoupled field
mode, in the same short-wavelength limit. Taking the time derivative of
(\ref{deB2D}), and using (\ref{S2D}), yields%
\begin{equation}
a^{3}\ddot{q}_{\mathbf{k}1}=-\frac{\partial}{\partial q_{\mathbf{k}1}%
}(V+Q),\ \ \ a^{3}\ddot{q}_{\mathbf{k}2}=-\frac{\partial}{\partial
q_{\mathbf{k}2}}(V+Q) \label{B2D}%
\end{equation}
(with $a^{3}\approx\mathrm{const.}$), where%
\[
V=\frac{1}{2}ak^{2}\left(  q_{\mathbf{k}1}^{2}+q_{\mathbf{k}2}^{2}\right)
\]
is the classical potential and%
\[
Q=-\frac{1}{2a^{3}}\frac{\nabla_{1,2}^{2}|\psi_{\mathbf{k}}|}{|\psi
_{\mathbf{k}}|}%
\]
is the quantum potential (with $\nabla_{1,2}^{2}\equiv\partial^{2}/\partial
q_{\mathbf{k}1}^{2}+\partial^{2}/\partial q_{\mathbf{k}2}^{2}$).

In Bohm's dynamics for this field system, (\ref{B2D}) are the equations of
motion, while (\ref{deB2D}) are initial equilibrium conditions only, just as
in the low-energy particle theory.

It is now readily seen that if we consider Bohm's dynamics for an initial
nonequilibrium state, with%
\[
\dot{q}_{\mathbf{k}1}\neq\frac{1}{a^{3}}\frac{\partial s_{\mathbf{k}}%
}{\partial q_{\mathbf{k}1}},\ \ \ \ \dot{q}_{\mathbf{k}2}\neq\frac{1}{a^{3}%
}\frac{\partial s_{\mathbf{k}}}{\partial q_{\mathbf{k}2}}%
\]
at some initial time $t_{i}$, then the field amplitudes $(q_{\mathbf{k}%
1},q_{\mathbf{k}2})$ will show the same instability as we found for the
low-energy particle case. For example, if the initial wave function is a
superposition of the ground state plus a few of the first excited states, then
as $|\psi_{\mathbf{k}}(q_{\mathbf{k}1},q_{\mathbf{k}2},t)|^{2}$ evolves it
will remain localised around the origin $(q_{\mathbf{k}1},q_{\mathbf{k}%
2})=(0,0)$. Whereas, because trajectories $(q_{\mathbf{k}1}(t),q_{\mathbf{k}%
2}(t))$ can leave this localised region and move off to infinity, the actual
ensemble distribution $\rho_{\mathbf{k}}(q_{\mathbf{k}1},q_{\mathbf{k}2},t)$
can evolve far away from equilibrium, just as we saw for the particle case.

The physical consequences of such behaviour for the field would be quite
drastic. If the field amplitudes $|q_{\mathbf{k}r}|$ grow unboundedly large,
this means that the field $\phi(\mathbf{x},t)$ itself becomes unbounded in
magnitude. Similar results would be obtained for the electromagnetic field,
for example, resulting in unboundedly large electric and magnetic field
strengths even in the vacuum. This is grossly at variance with observation.

\section{Conclusion}

We have shown that Bohm's dynamics is unstable. Small deviations from initial
equilibrium do not relax and instead grow with time.

In de Broglie's dynamics, conservation of the configuration-space distribution
(\ref{Born}) implies that it is an equilibrium state. In Bohm's dynamics,
conservation of the phase-space distribution (\ref{equ1}) similarly implies
that it is an equilibrium state. Our analysis shows that, despite this prima
facie similarity, there is a fundamental difference: in de Broglie's dynamics
the equilibrium state is stable whereas in Bohm's dynamics it is not.

On the basis of these results we conclude that Bohm's dynamics (as we have
defined it) is untenable as a physical theory. It agrees with quantum theory
and with observation only if very special initial conditions are assumed.
Specifically, the initial momentum distribution in phase space must be
concentrated exactly on the surface defined by $p=\nabla_{q}S$.

If Bohm's dynamics were correct it would be unreasonable to expect to see an
effective quantum theory today -- even approximately -- in contradiction with
observation. This is in sharp contrast with de Broglie's dynamics, where
efficient relaxation to equilibrium implies that one should expect to see
equilibrium at later times (except, possibly, for very long-wavelength modes
in the early universe (Valentini 2007, 2008b, 2010; Colin and Valentini
2013)). It is then reasonable to conclude that, while de Broglie's dynamics is
a viable physical theory, Bohm's dynamics is not.

To avoid this conclusion, there are two possible responses, each of which
seems unconvincing:

(i) It might be asserted that the extended quantum equilibrium state
(\ref{equ1}) is `absolute', in the sense of defining a preferred measure of
`typicality' for the initial conditions of the universe. An analogous claim
has been made by some workers for the standard quantum equilibrium state
(\ref{Born}) in the context of de Broglie's dynamics (D\"{u}rr, Goldstein and
Zangh\`{\i} 1992). This last approach has been criticised on grounds of
circularity (Valentini 1996, 2001). But even leaving such criticisms aside,
for the case of Bohm's dynamics how could one justify using a particular
measure of typicality when there are two equilibrium distributions
(\ref{equ1}) and (\ref{equ2})? The `wrong' choice -- the equilibrium measure
(\ref{equ2}) -- would conflict grossly with observation. And yet there appears
to be no \textit{a priori} reason to prefer (\ref{equ1}) over (\ref{equ2}) as
a probability (or typicality) measure for initial conditions.

(ii) It might be suggested that Bohm's dynamics is only an approximation, and
that corrections from a deeper theory will (in reasonable circumstances) drive
the phase-space distribution to equilibrium. Such a suggestion was in fact
made by Bohm (1952a, p. 179). While this may turn out to be the case, the fact
remains that Bohm's dynamics as it stands is unstable and therefore (we claim) untenable.

The results of this paper highlight the importance of stability as a criterion
for hidden-variables theories to be acceptable.

In our view Bohm's 1952 Newtonian reformulation of de Broglie's 1927
pilot-wave dynamics was a mistake, and we ought to regard de Broglie's
original formulation as the correct one. Such a preference is no longer merely
a matter of taste: we have presented concrete physical reasons for preferring
de Broglie's dynamics over Bohm's.

\textbf{Acknowledgements.} We are grateful to Ward Struyve for participating
in discussions during the course of this work. SC carried out part of this
work at the Centre for Quantum Dynamics, Griffith University, Australia. This
research was funded in part by the John Templeton Foundation.

\begin{center}
\textbf{References}
\end{center}

Bacciagaluppi, G. and Valentini, A. 2009 \textit{Quantum theory at the
crossroads: reconsidering the 1927 Solvay conference}. Cambridge: Cambridge
University Press. [arXiv:quant-ph/0609184]

Bennett, A. F. 2010 Relative dispersion and quantum thermal equilibrium in de
Broglie-Bohm mechanics. \textit{J. Phys. A: Math. Theor.} \textbf{43}, 195304.

Bohm, D. 1952a A suggested interpretation of the quantum theory in terms of
`hidden' variables. I. \textit{Phys. Rev.} \textbf{85}, 166.

Bohm, D. 1952b A suggested interpretation of the quantum theory in terms of
`hidden' variables. II. \textit{Phys. Rev.} \textbf{85}, 180.

Colin, S. 2012 Relaxation to quantum equilibrium for Dirac fermions in the de
Broglie-Bohm pilot-wave theory. \textit{Proc. Roy. Soc. Lond. A} \textbf{468},
1116. [arXiv:1108.5496]

Colin, S. and Valentini, A. 2013 Mechanism for the suppression of quantum
noise at large scales on expanding space. \textit{Phys. Rev. D} \textbf{88},
103515. [arXiv:1306.1579]

Davies, P. C. W. 1974 \textit{The physics of time asymmetry}. Berkeley:
University of California Press.

de Broglie, L. 1928 La nouvelle dynamique des quanta, in:
\textit{\'{E}lectrons et photons: rapports et discussions du cinqui\`{e}me
conseil de physique}. Paris: Gauthier-Villars. [English translation in:
Bacciagaluppi and Valentini (2009).]

D\"{u}rr, D., Goldstein, S. and Zangh\`{\i}, N. 1992 Quantum equilibrium and
the origin of absolute uncertainty. \textit{J. Stat. Phys.} \textbf{67}, 843. [arXiv:quant-ph/0308039]

Efthymiopoulos, C. and Contopoulos, G. 2006 Chaos in Bohmian quantum
mechanics. \textit{J. Phys. A: Math. Gen.} \textbf{39}, 1819.

Goldstein, S. and Struyve, W. 2014 On quantum potential dynamics. arXiv:1312.1990v2.

Pearle, P. and Valentini, A. 2006 Quantum mechanics: generalizations, in:
\textit{Encyclopaedia of mathematical physics}, eds. J.-P. Fran\c{c}oise
\textit{et al}.. North-Holland: Elsevier. [arXiv:quant-ph/0506115]

Towler, M. D., Russell, N. J. and Valentini, A. 2012 Time scales for dynamical
relaxation to the Born rule. \textit{Proc. Roy. Soc. Lond. A} \textbf{468},
990. [arXiv:1103.1589]

Valentini, A. 1991a Signal-locality, uncertainty, and the subquantum
\textit{H}-theorem. I. \textit{Phys. Lett. A} \textbf{156}, 5.

Valentini, A. 1991b Signal-locality, uncertainty, and the subquantum
\textit{H}-theorem. II. \textit{Phys. Lett. A} \textbf{158}, 1.

Valentini, A. 1992 On the pilot-wave theory of classical, quantum and
subquantum physics. PhD thesis, International school for advanced studies,
Trieste, Italy. [https://digitallibrary.sissa.it/bitstream/handle/1963/5424/PhD\_Valentini\_Antony.pdf?sequence=1]

Valentini, A. 1996 Pilot-wave theory of fields, gravitation and cosmology, in:
\textit{Bohmian mechanics and quantum theory: an appraisal}, eds. J. T.
Cushing \textit{et al}.. Dordrecht: Kluwer.

Valentini, A. 2001 Hidden variables, statistical mechanics and the early
universe, in: \textit{Chance in physics: foundations and perspectives}, eds.
J. Bricmont \textit{et al}.. Berlin: Springer. [arXiv:quant-ph/0104067]

Valentini, A. 2002 Subquantum information and computation. \textit{Pramana --
J. Phys.} \textbf{59}, 269. [arXiv:quant-ph/0203049]

Valentini, A. 2007 Astrophysical and cosmological tests of quantum theory.
\textit{J. Phys. A: Math. Theor.} \textbf{40}, 3285. [arXiv:hep-th/0610032]

Valentini, A. 2008a Hidden variables and the large-scale structure of
space-time, in: \textit{Einstein, relativity and absolute simultaneity}, eds.
W. L. Craig and Q. Smith. London: Routledge. [arXiv:quant-ph/0504011]

Valentini, A. 2008b De Broglie-Bohm prediction of quantum violations for
cosmological super-Hubble modes. arXiv:0804.4656.

Valentini, A. 2009 Beyond the quantum. \textit{Physics World} \textbf{22N11},
32. [arXiv:1001.2758]

Valentini, A. 2010 Inflationary cosmology as a probe of primordial quantum
mechanics. \textit{Phys. Rev. D} \textbf{82}, 063513. [arXiv:0805.0163]

Valentini, A. and Westman, H. 2005 Dynamical origin of quantum probabilities.
\textit{Proc. Roy. Soc. Lond. A} \textbf{461}, 253. [arXiv:quant-ph/0403034]

\end{document}